\documentclass[]{elsart}

\usepackage{graphicx}% Include figure files
\usepackage{dcolumn}% Align table columns on decimal point
\usepackage{amsmath}

\newcommand{\cer}{\v{C}erenkov~}
\newcommand{\cas}{$\Xi^-$~}
\newcommand{\sig}{$\Sigma^+$\ }
\newcommand{\lam}{$\Lambda^0$\ }

\newcommand{\cpipi}{$\Xi_c^+ \to \Xi^- \pi^+ \pi^+$\ }
\newcommand{\skpi}{$\Xi_c^+\to \Sigma^+K^-\pi^+$\ }
\newcommand{\pkpi}{$\Xi_c^+\to pK^-\pi^+$\ }
\newcommand{\lkpp}{$\Xi_c^+\to \Lambda^0K^-\pi^+\pi^+$\ }
\newcommand{\cc}{$\Xi_c^+$\ }
\newcommand{\myalpha}{4}
\begin{document}

\begin{frontmatter}
\title{A New Measurement of the \cc Lifetime}

The FOCUS Collaboration

%%%%%%% Do not change authors here.  Use the database!
\author[ucd]{J.~M.~Link},
\author[ucd]{M.~Reyes},
\author[ucd]{P.~M.~Yager},
\author[cbpf]{J.~C.~Anjos},
\author[cbpf]{I.~Bediaga},
\author[cbpf]{C.~G\"obel},
\author[cbpf]{J.~Magnin},
\author[cbpf]{A.~Massafferri},
\author[cbpf]{J.~M.~de~Miranda},
\author[cbpf]{I.~M.~Pepe},
% ...

\address[ucd]{University of California, Davis, CA 95616} 
\address[cbpf]{Centro Brasileiro de Pesquisas F\'isicas, Rio de Janeiro, RJ, Brasil} 
\address[cinv]{CINVESTAV, 07000 M\'exico City, DF, Mexico} 
\address[cu]{University of Colorado, Boulder, CO 80309} 
\address[fnal]{Fermi National Accelerator Laboratory, Batavia, IL 60510} 
\address[fras]{Laboratori Nazionali di Frascati dell'INFN, Frascati, Italy I-00044} 
\address[ui]{University of Illinois, Urbana-Champaign, IL 61801} 
\address[iu]{Indiana University, Bloomington, IN 47405} 
\address[korea]{Korea University, Seoul, Korea 136-701} 
\address[milan]{INFN and University of Milano, Milano, Italy} 
\address[nc]{University of North Carolina, Asheville, NC 28804} 
\address[pavia]{Dipartimento di Fisica Nucleare e Teorica and INFN, Pavia, Italy} 
\address[pr]{University of Puerto Rico, Mayaguez, PR 00681} 
\address[sc]{University of South Carolina, Columbia, SC 29208} 
\address[ut]{University of Tennessee, Knoxville, TN 37996} 
\address[vu]{Vanderbilt University, Nashville, TN 37235} 
\address[wisc]{University of Wisconsin, Madison, WI 53706} 

\date{\today}

\begin{abstract}
{A precise determination of the charm-strange baryon \cc lifetime is presented. 
The data were accumulated by the Fermilab high-energy photoproduction 
experiment \mbox{FOCUS}.
The measurement is made with  
300 \cpipi decays, 130 \skpi decays, 45  \pkpi decays and 58 \lkpp decays.
The \cc lifetime is measured to be 0.439$\pm$0.022$\pm$0.009 ps.}
\end{abstract}
%%\PACS 13.30.Eg 14.20.Lb
\end{frontmatter}

\section{Introduction}
The lifetime hierarchy of the weakly decaying charm mesons is well established
\cite{Groom:2000in}.
However, the pattern of the predicted lifetimes for the weakly decaying charm baryons
agrees only qualitatively with experimental results 
\cite{Guberina:1997yx,Bellini:1997ra}.
The $\Lambda^+_c$ lifetime is known to an accuracy of $\sim$5\%
\cite{Mahmood:2000tw,Kushnirenko:2000ed}, but the others, ($\Xi^+_c$, $\Xi^0_c$,
$\Omega^0_c$), have uncertainties on the order of 20\%. 
Interestingly, baryon sector lifetime measurements provide information on quark
interference and W-exchange.
The essential difference from the mesons is that W-exchange among the valence quarks of
the baryon is neither color nor helicity suppressed.
The measured \cc lifetime is larger than theory predicts, but there is a large
experimental uncertainty \cite{Frabetti:1998kr,Frabetti:1993yy}.
A more precise measurement could be conclusive in testing predictions in this sector.

In the FOCUS spectrometer\footnote{FOCUS spectrometer is an upgraded version of the
Fermilab E687 spectrometer~\cite{Frabetti:1992au}.}
high energy photons with $\langle E \rangle \approx 180$~ GeV interact in a segmented
BeO target to produce charmed particles.
Charged particles are tracked in the target region by two silicon vertex
detectors which provide excellent vertex separation between the production and decay
vertices.
The average proper time resolution is $\approx 50$ fs for the modes used in this
analysis.
Downstream tracking and momentum measurement is performed by a system of five
multiwire proportional chambers (MWPC) and two magnets of opposite polarity. 
The upstream magnet (M1) is positioned downstream of the silicon detectors and
in front of the MWPC system. The second magnet (M2) lies between the third and
fourth MWPC stations.
Charged particle identification is provided by three multicell threshold \cer
counters and two muon systems. 
One hadronic and two electromagnetic calorimeters are used to measure
particle energy.
%
%%%%%%%%%%%%%%%%%%%%%% Cc --> Hyperon description
%
\section{Reconstruction of hyperons $\Xi^-$ and $\Sigma^+$}
A detailed description of the \cas and \sig reconstruction can be found 
elsewhere \cite{Link:2001jc}.
The \cas decays into $\Lambda^0 \pi^-$, with the \lam being
reconstructed, if possible, through the decay $\Lambda^0 \to
p\pi^-$.\footnote{Throughout this paper, the charge conjugate state is implied whenever
the decay mode of a particular charge is stated.}
The $\Xi^-$'s are reconstructed according to where and how the decay occurs.
We consider four categories: 
Type1 where the \cas decays promptly without leaving a track in the silicon detectors; 
Type2 where the \cas decays after passing through the silicon and the \lam daughter is
fully reconstructed; 
MV decays, which are topologically identical to Type2 decays, but are reconstructed
from the intersection of the three MWPC tracks from the \cas decay with the \cas 
silicon track; 
Kink decays where the \cas passes through the silicon and the \lam daughter is not
reconstructed.
If a \cas candidate is reconstructed both as Type2 and MV, then we choose the Type2 to
avoid duplication. In Kink type decays, due to the failure to reconstruct the
$\Lambda^0$, there is normally a two-fold ambiguity in the determination of the
momentum. However, by requiring that the decay occurs in M1 we remove this ambiguity. 
Our data is composed of 9\% Type1, 66\% Type2, 6\% MV and 19\% Kink. We have more than
one million reconstructed $\Xi^-$'s in all four types. 

The \sig baryon can decay into $p\pi^0$ or $n\pi^+$. Each channel is
studied separately. The reconstruction is similar to the Kink category described above.
For the neutron in the $\Sigma^+\to n\pi^+$ decay we require $0.3<E/p<2.0$, where $E$
is the energy deposited in the calorimeters and $p$ is the reconstructed momentum.
For the $\Sigma^+\to p\pi^0$ decays, the protons must be positively identified in the
\cer
system. As in the case of Kink type $\Xi^-$'s, we have a two-fold ambiguity in the 
determination of the \sig momentum. This ambiguity can be partially removed for
the $\Sigma^+\to n\pi^+$ decay by using the location of energy deposition in the
calorimeters.
%
%%%%%%%%%%%%%%%%%%%% Cascade_c reconstruction.
%
\section{Reconstruction of \cc candidates}
\begin{figure*}[htb]
  \begin{center}
     \includegraphics[width=1.3in]{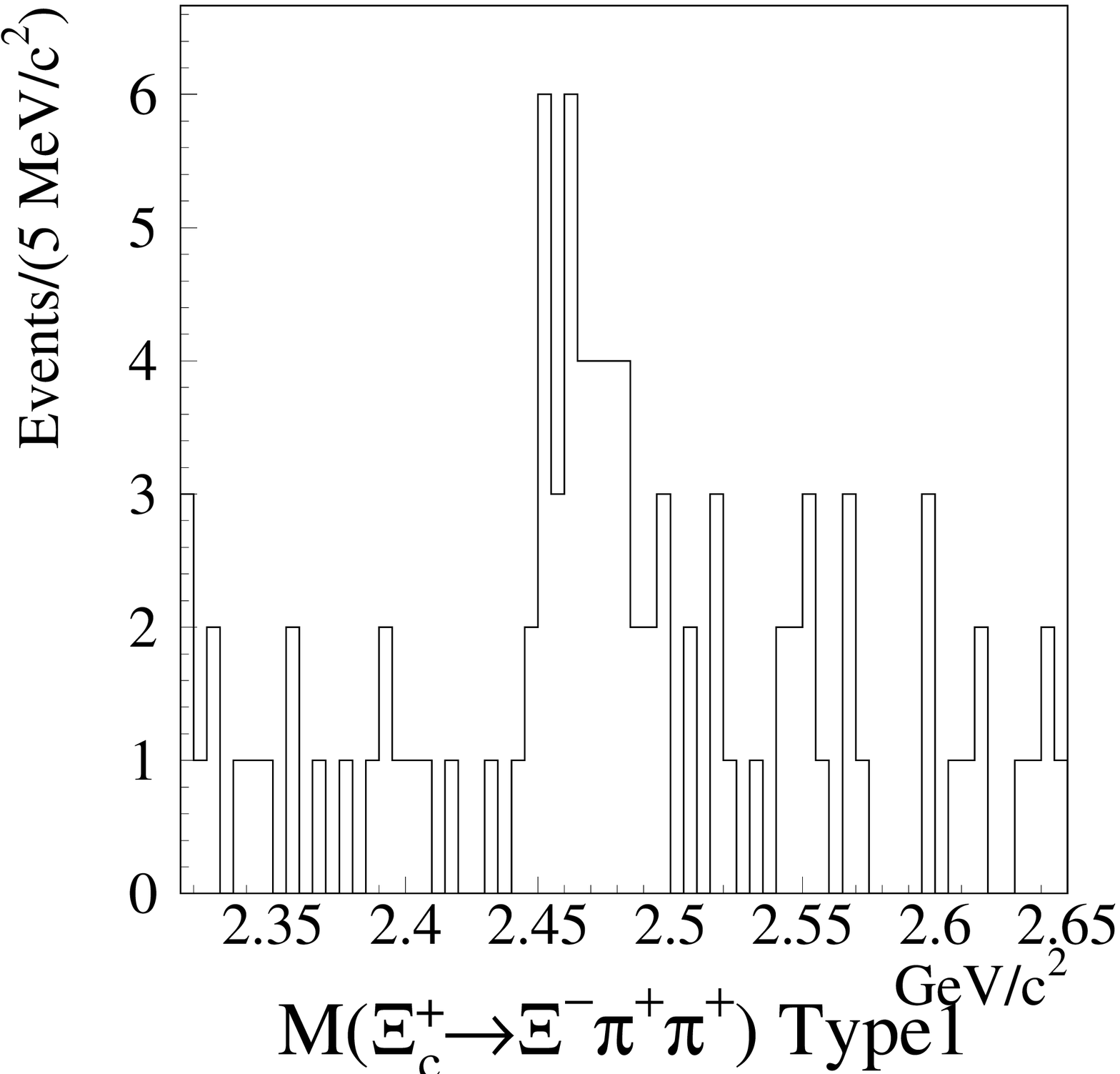}
     \includegraphics[width=1.3in]{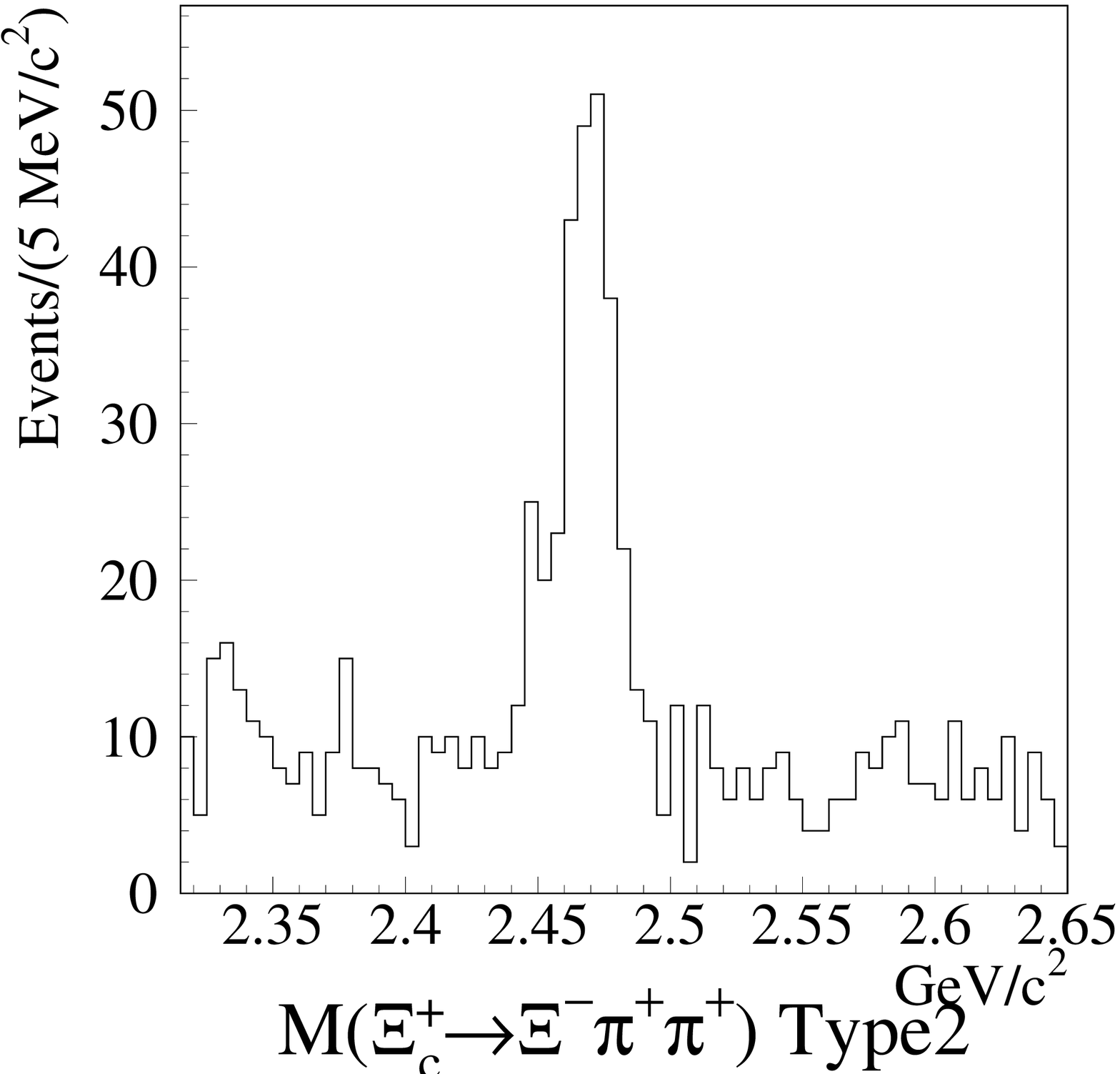}
     \includegraphics[width=1.3in]{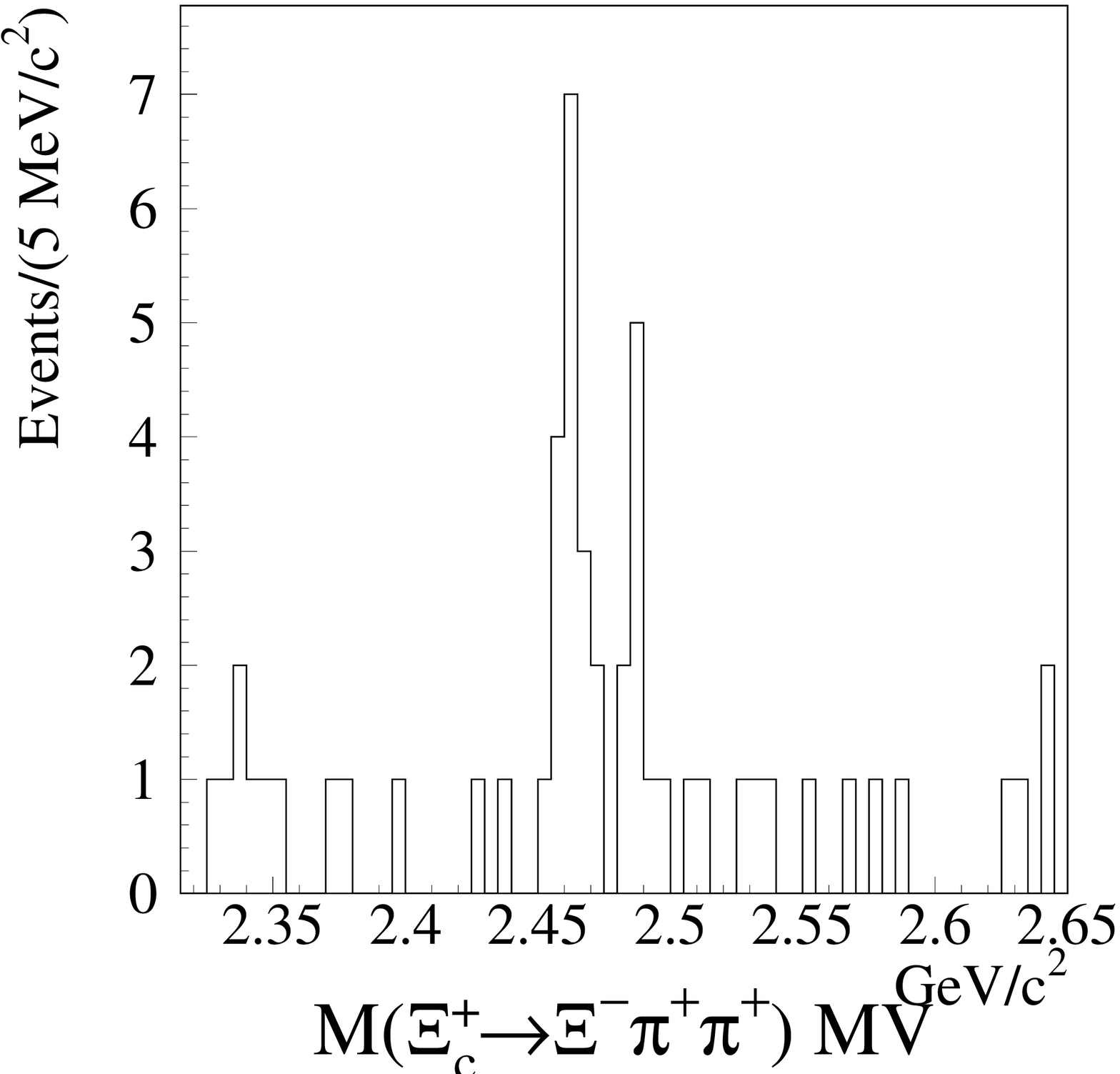}
     \includegraphics[width=1.3in]{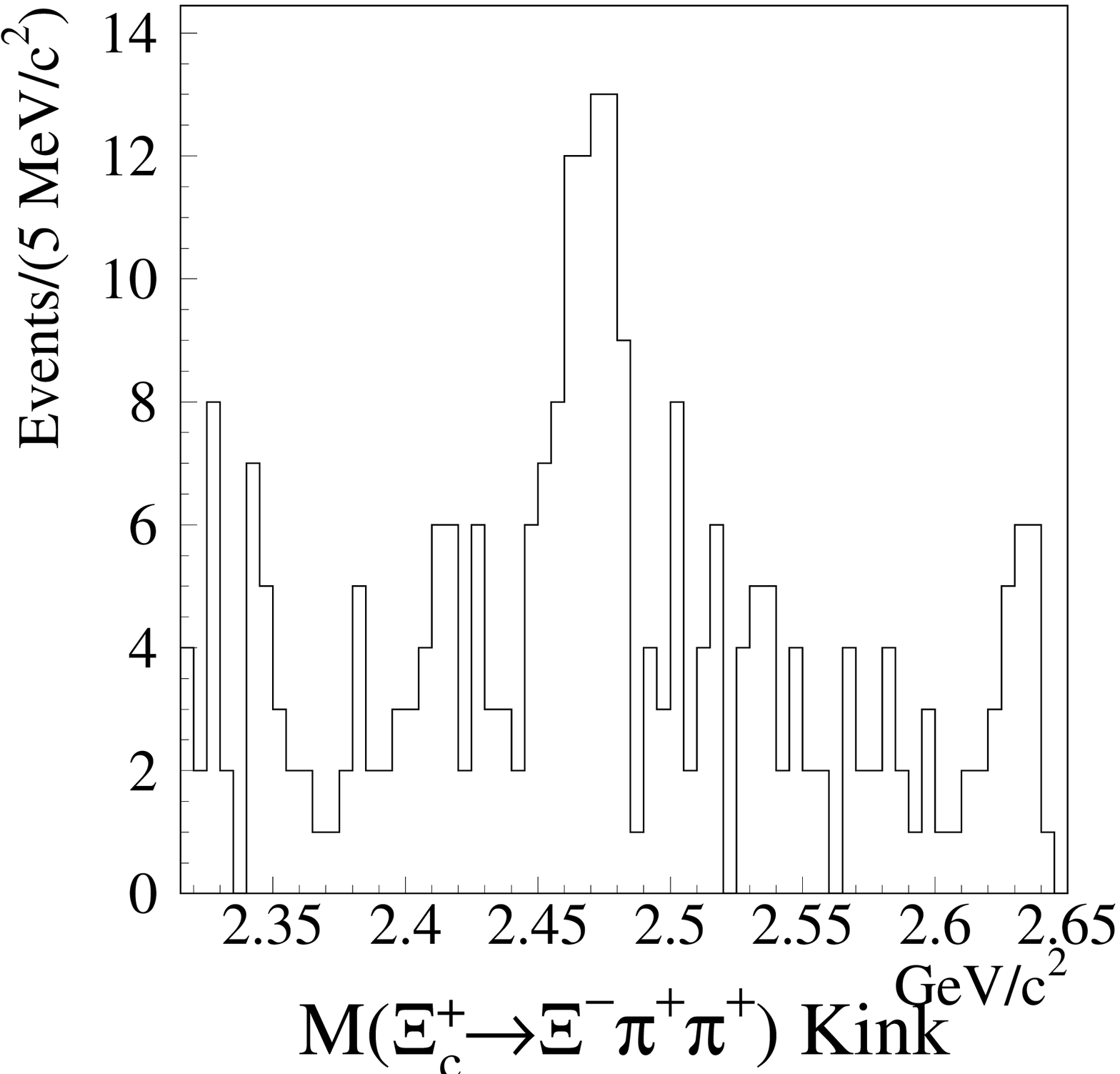}\\
     \includegraphics[width=1.3in]{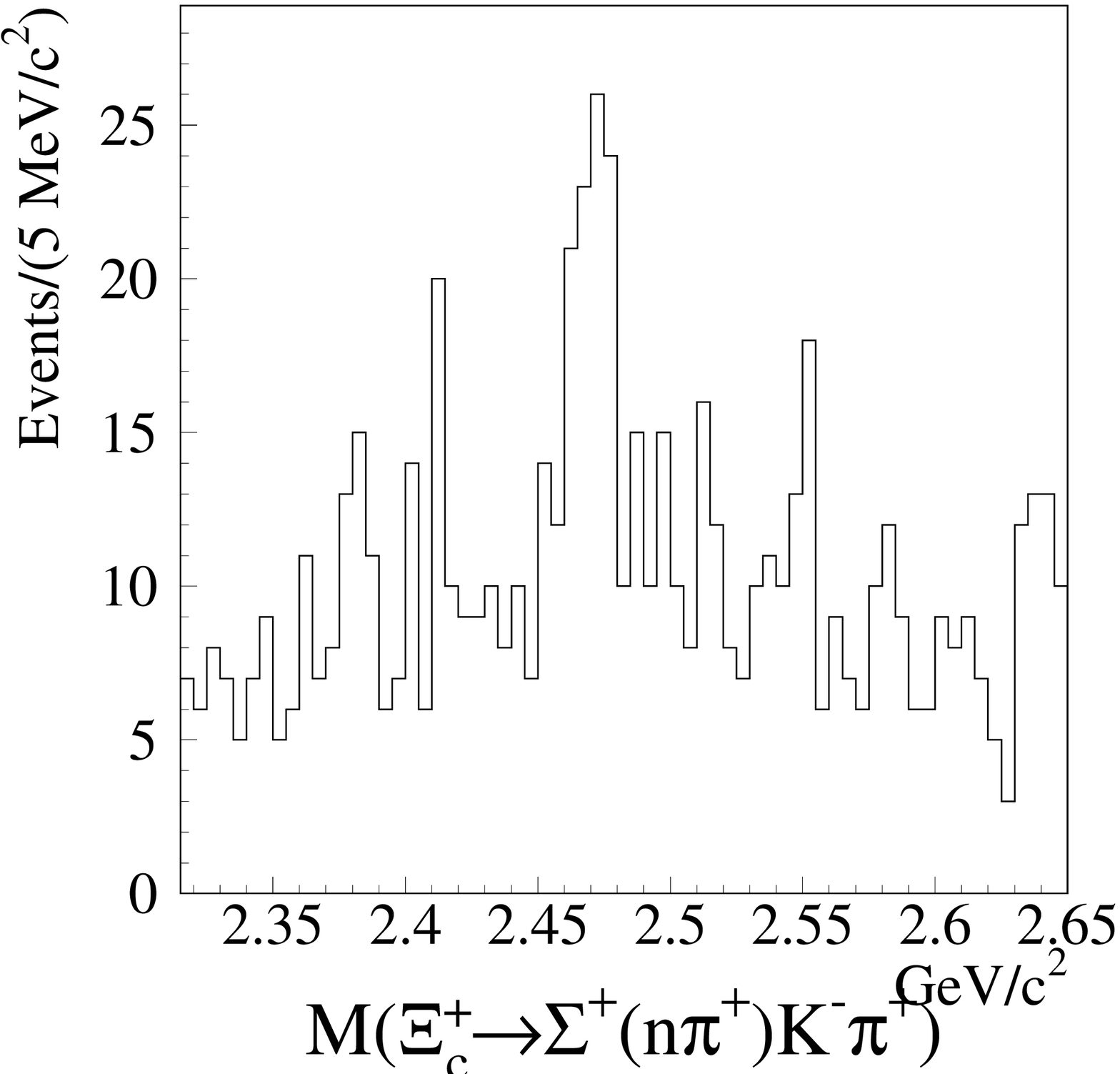}
     \includegraphics[width=1.3in]{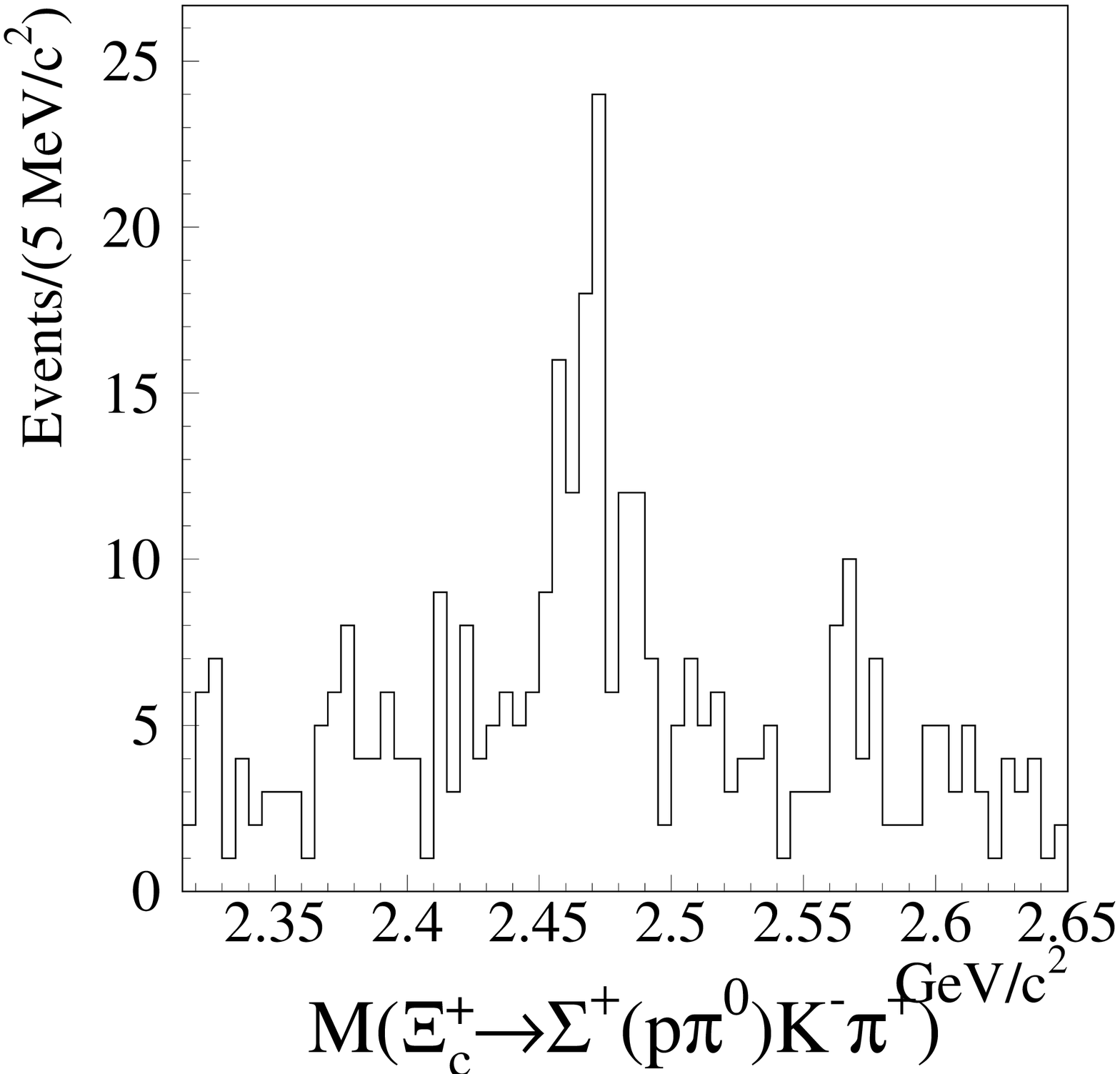}
     \includegraphics[width=1.3in]{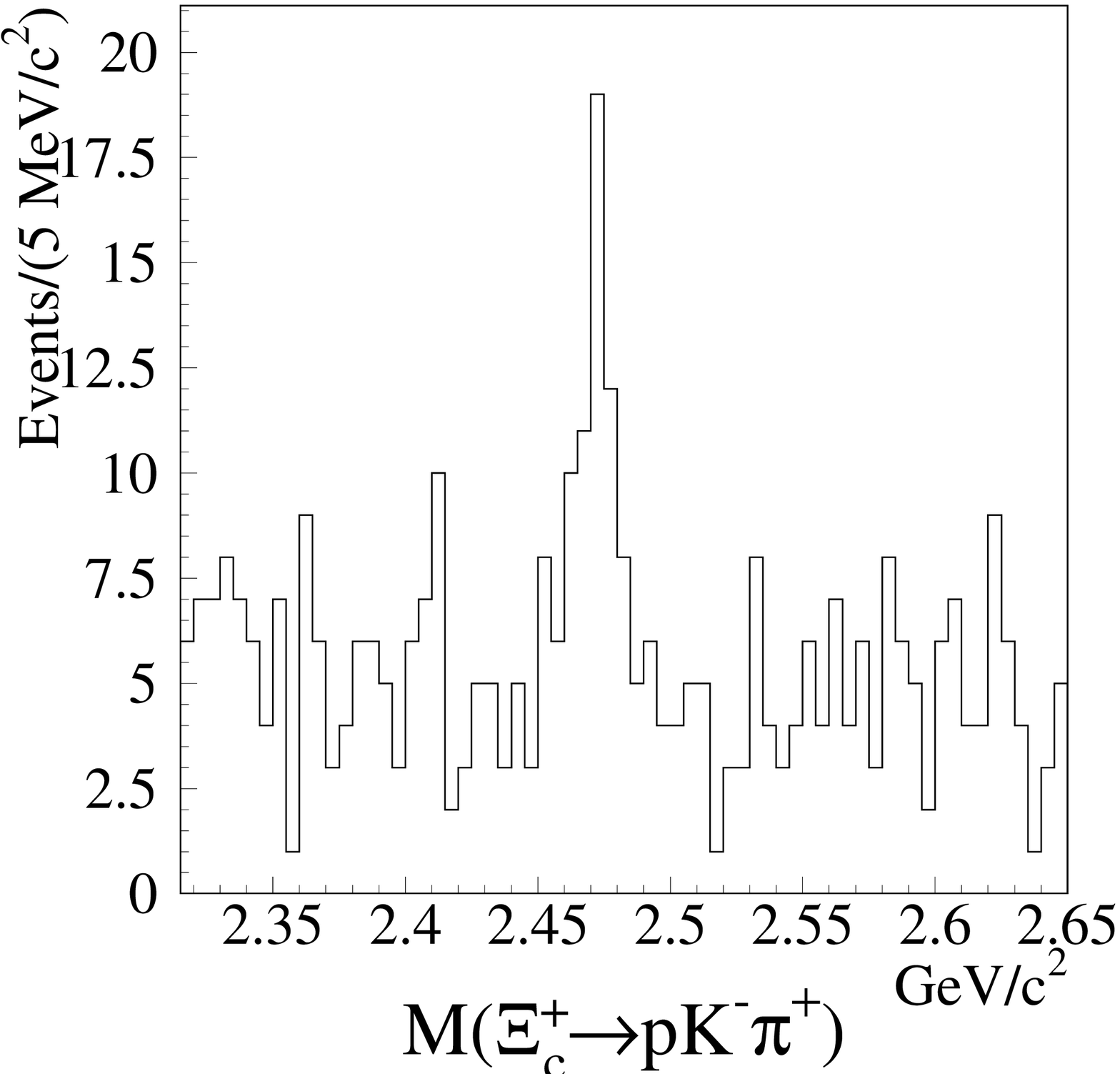}
     \includegraphics[width=1.3in]{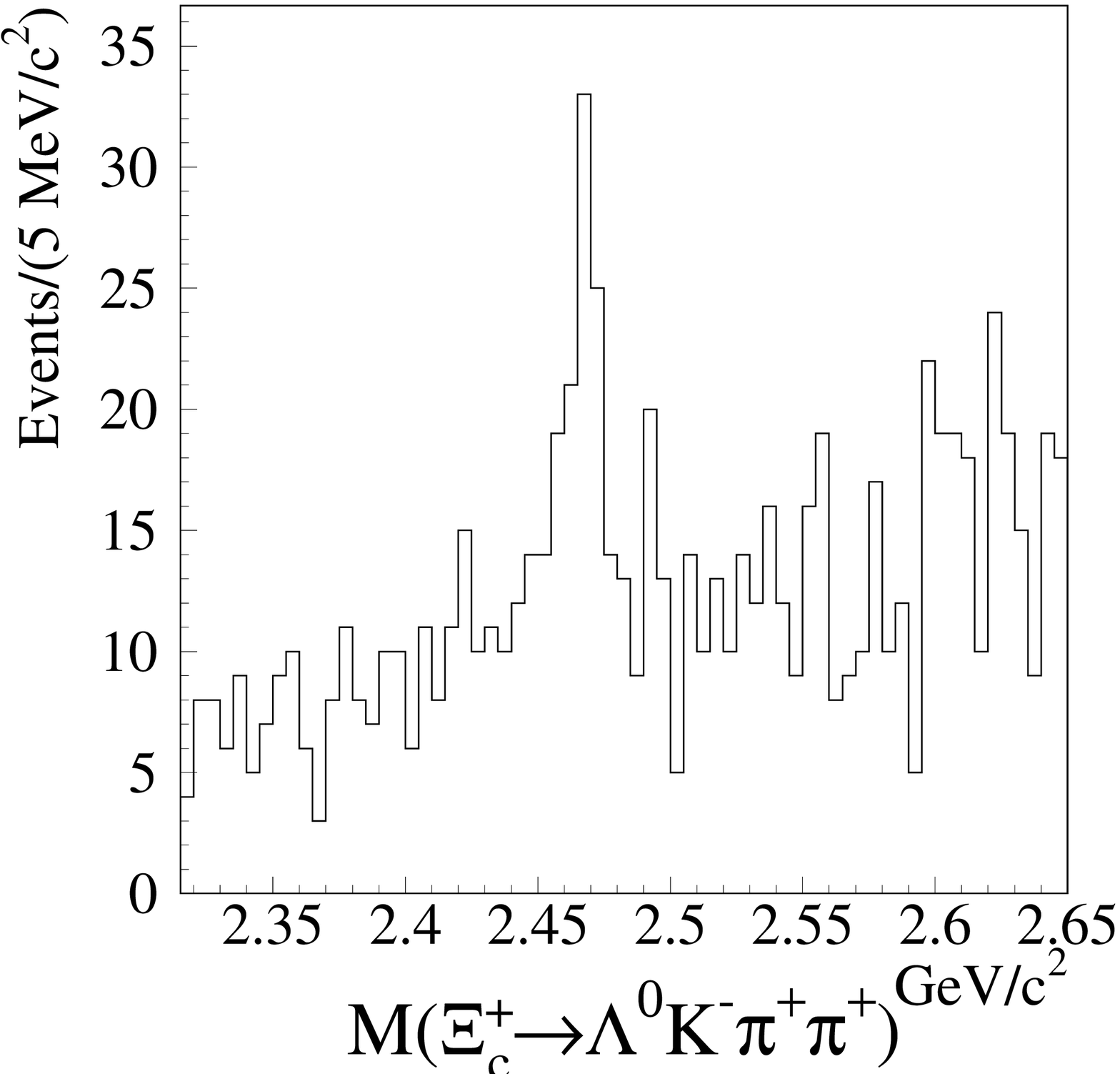}
  \caption{$\Xi^- \pi^+ \pi^+$ (4 reconstruction methods), 
  $\Sigma^+ K^- \pi^+$ (2 decay modes),
  $p K^-\pi^+$ and 
  $\Lambda^0 K^- \pi^+ \pi^+$ 
  invariant mass distributions with cuts as described in the text.}
 \label{fig:xc}
 \end{center}
\end{figure*}

The \cc candidate is reconstructed using a candidate driven vertexing
algorithm \cite{Frabetti:1992au}. A
\cc candidate vertex is formed using the silicon track information  of the decay
daughters when available.
This (secondary) vertex is required to have a confidence level (CL) above a value
optimized for each topology reconstructed.
We construct a seed track using the momentum vector of the \cc candidate and
intersect it with at least two other tracks to form a
primary vertex. This primary vertex is required to have a confidence level greater than
1\% and to be in the target material to within 3 units of the error in the primary
vertex position (TGM$<$3). We tighten this requirement to TGM$<$0 for the cases where
the \cas or \sig are partially reconstructed.
Other cuts used to optimize the signal are the confidence level that any other
track originates from the secondary vertex (ISO2), the error in the proper time
($\sigma_{t}$), and the significance of separation of the primary and secondary
vertices ($L/\sigma_L$).
\cer particle identification is accomplished by constructing $\chi^2$ like variables
for the different particle hypotheses \cite{Link:2001pg}.
Briefly, we compute likelihoods for the various stable particle hypotheses
$e$, $\pi$, $K$ and $p$. The pion consistency of a track is defined by a requirement on
$\Delta W_{\pi}=W_{\text{{min}}}-W_{\pi}$, where $W_{\pi}$ is the negative log 
likelihood of the pion hypothesis and $W_{\text{{min}}}$ is the minimum negative log
likelihood of the other three hypotheses.
Similarly, we define $\Delta W_K= W_{\pi}-W_K$ and $\Delta W_p= W_{\pi}-W_p$ for use in
identifying kaons and protons.
We require $\Delta W_{\pi}>$-6 and $\Delta W_K>$2 for pion and kaon identification
in all modes.
Additional particle identification is mode dependent and is described below.
The resulting invariant mass distributions are shown in Figure \ref{fig:xc}; the
combined plot is shown in Figure \ref{fig:ltime}(a).
%
%%%%%%%%%%%%%%%%%%%% Cascade_c --> cascadepipi description.
%
\subsection{\cpipi selection criteria}
The secondary vertex is required to have a CL greater than 2\%
except for Type1 \cas where we require it to be greater than 10\%.
For Kink type decays we require that $\sigma_{t}$ be less than 0.06 ps and
that ISO2 be less than 0.01\%.
The additional cuts are necessary since we cannot make an invariant mass cut on \cas
candidates of this type. We demand that $L/\sigma_L$ be greater than 4.
%
%%%%%%%%%%%%%%%%%%%% Cascade_c --> SigmaKpi description.
%
\subsection{\skpi selection criteria}
If there is an ambiguity in the $\Sigma^+$ momentum determination we resolve it as
follows. First, we select the higher momentum solution when the difference in 
the masses calculated using each solution is less than 30 MeV/$c^2$. 
When the difference is greater than
30 MeV/$c^2$ we reject a solution candidate if the other \skpi mass is within 
2.5$\sigma$ of the nominal \cc mass. For a small fraction of cases this can lead
to rejecting or accepting both solutions. Resolving the ambiguity in this way
may create a bias and we have studied this as a possible source of systematic 
uncertainty.
The CL of the secondary vertex must be greater than $2\%$ and we require 
$\sigma_{t}$ to be less than 0.07 ps. We ask that $L/\sigma_L$ be greater than 7.
%
%%%%%%%%%%%%%%%%%%%% Cascade_c --> pKpi description.
%
\subsection{\pkpi selection criteria}
We identify protons by requiring that $\Delta W_p>10$
and that the proton hypothesis is favored over the kaon hypothesis by two units of
likelihood i.e. $W_{K}-W_p>2$.  
We demand that  $W_p-W_{K}>0$ for the kaon candidate.
For pion candidates we require the momentum to be greater than 5 GeV/$c$.
The secondary vertex must have a CL greater than 15\% and  ISO2 less than 0.01\%. 
To reduce combinatorial background, the \cc candidate momentum is required to be less
than 120 GeV/$c$.
Long lifetime backgrounds from charmed mesons, where one particle is misidentified, are
removed by making an invariant mass cut. In this way we eliminate contamination from
the decays $D^+ \to K^-\pi^+\pi^+$, $D^+(D_s^+) \to
K^-K^+\pi^+$ and $D^0 \to K^+K^-$.
Finally, cuts of $\sigma_{t}$ less than 0.08 ps and $L/\sigma_L$ greater than 7 are
applied.

%%%%%%%%%%%%%%%%%%%% Cascade_c --> lambda K pi pi description.
%
\subsection{\lkpp selection criteria}
We select \lam candidates which decay downstream of the silicon vertex detector.
To reduce contamination from $K_S \to \pi^+\pi^-$ decays, we require $\Delta W_p>8$ for
the proton in the \lam decay.
The secondary vertex must have a CL greater than 15\% and ISO2 less than 0.01\%. 
Cuts of $\sigma_{t}$ less than 0.10 ps and $L/\sigma_L$ greater than 4
are applied.

%%%%%%%%%%%%%%%%%%%% Lifetime Technique.
%
\section{Lifetime technique.}
\begin{figure}[ht]
  \begin{center}
     \includegraphics[width=3in]{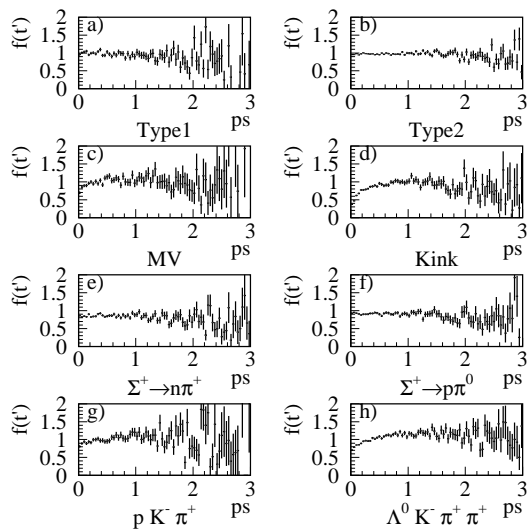}
  \caption{Lifetime
  correction functions for the different decay modes and topologies described earlier.
   }
  \label{fig:ft}
 \end{center}
\end{figure}

We perform a binned maximum likelihood fit \cite{Frabetti:1991nd} to extract the
lifetime of the \cc from the aforementioned decay channels.
We fit the reduced proper time ($t'$) distribution, defined as
$t'=(L-N\sigma_L)/\beta\gamma c$ where $N$ is the vertex detachment
cut, $\beta c$ is the particle velocity and $\gamma$ is the Lorentz boost factor
to the $\Xi_c^+$ center of mass frame. 
The $t'$ distribution for $\Xi_c^+$ is of the form $e^{-t'/\tau}$, where $\tau$ is the
lifetime of the $\Xi_c^+$. 
A fit is made to the $t'$ distribution for events which lie within $\pm2\sigma$ of the
$\Xi_c^+$ mass, where $\sigma$ is $\sim$8 MeV$/c^2$ for the \lkpp channel and
$\sim$10 MeV$/c^2$ otherwise. 
The background is assumed to have the same lifetime behavior in the signal region and
in the sidebands which are 4--12$\sigma$ away from the peak.
Taking $S$ as the number of signal events in the mass region and $B$ as the total
number of background events in the same region, the expected number of events $n_i$ in
the $i^{th}$ reduced proper time bin centered at $t'$ is given by:
\begin{equation} 
n_{i} = S\frac{f(t'_{i})e^{-t'_{i}/\tau}}
                 { \displaystyle \sum_{ i}^{} f(t'_{i})e^{-t'_{i}/\tau}  }+
           B\frac{b_{i}}{\displaystyle \sum_{i}^{} b_{i}}
\label{eq:ni}
\end{equation}
where $b_i$ describes the background reduced proper time evolution as
estimated from sidebands and $f(t'_{i})$ is a correction function,
which takes into account the effects of spectrometer acceptance, analysis cut
efficiencies, and absorption of the particles as a function of the reduced
proper time. 
In Figure \ref{fig:ft} we plot the correction function for each decay mode.
The large corrections in some modes at low reduced proper time are due to the
suppression of short lived decays by our selection cuts.
The likelihood is constructed from the product of the Poisson probability of observing
$s_i$ events when $n_i$ are expected with the Poisson 
probability of observing $\sum b_{i}$ background events when {\myalpha}$B$
are expected. The factor of \myalpha\ accounts for the fact that the sideband region is
four times wider than the signal region.
The likelihood takes the form:
\begin{equation} 
{\mathcal L}=\displaystyle (\prod_{i} \frac{n_{i}^{s_{i}}e^{-n_{i}}}{s_{i}!})
  \times 
  (\frac{({\myalpha} B)^{\displaystyle \sum_{i}b_{i}}e^{-{\myalpha} B}}
  {(\displaystyle \sum_{i}b_{i})!})
\end{equation}
The combined likelihood
function is given by the product of the eight likelihoods as shown
in Eq.~(\ref{eq:hood}). 
\begin{equation}\begin{split}
 {\mathcal L}_{\Xi_c^+}={\mathcal L}^{\Xi^- \pi^+ \pi^+}_{\mathrm Type1}
  \times
 {\mathcal L}^{\Xi^- \pi^+ \pi^+}_{\mathrm Type2} 
  \times 
 {\mathcal L}^{\Xi^- \pi^+ \pi^+}_{\mathrm MV} 
  \times 
 {\mathcal L}^{\Xi^- \pi^+ \pi^+}_{\mathrm Kink}\\  
  \times
 {\mathcal L}^{\Sigma^+ K^- \pi^+}_{\Sigma^+(n\pi^+)}
  \times
 {\mathcal{L}}^{\Sigma^+ K^- \pi^+}_{\Sigma^+(p\pi^0)} 
  \times
 {\mathcal{L}}_{p K^- \pi^+}
  \times
 {\mathcal{L}}_{\Lambda^0 K^- \pi^+\pi^+}
\end{split}\label{eq:hood}
\end{equation}
There are nine parameters in the fit, one parameter for the lifetime $\tau$ and eight
parameters for the backgrounds one for each decay type.
Our result for the \cc lifetime with statistical errors is $0.435 \pm 0.022$ ps.

\section{Systematic Studies}
\begin{figure}
  \begin{center}
     \includegraphics[width=3.in]{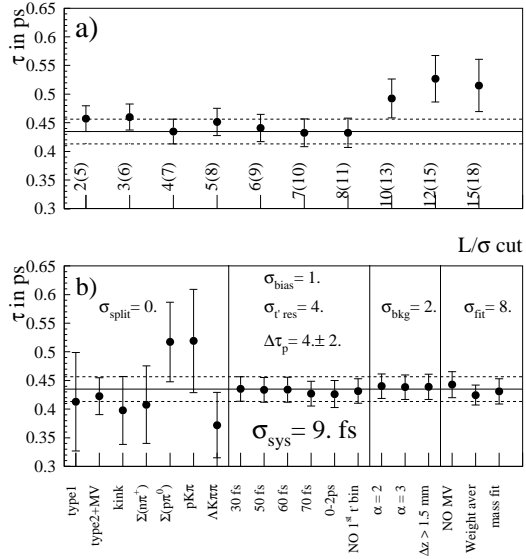}
  \caption{a) Lifetime stability versus $L/\sigma_L$; in parenthesis is the
$L/\sigma_L$ cut used for $\Sigma^+ K^- \pi^+$ and $p K^- \pi^+$.
  b) lifetime measurements for systematic
  studies.
  The solid line represent the central value with the dotted lines
  showing the extent of the statistical
  error at \mbox{$L/\sigma_L>4(7)$}.}
  \label{fig:lsevol}
 \end{center}
\end{figure}

We compute the lifetime using several different $L/\sigma_L$ cuts. The results are
shown in Figure \ref{fig:lsevol}(a).
Systematic effects were studied by computing the lifetimes of data samples split
by individual \cas topologies and modes. 
The results are shown in Figure \ref{fig:lsevol}(b).
All variations are consistent within statistical uncertainties and do not contribute
to a systematic uncertainty. 
We have investigated systematic effects due to the $t'$ resolution by examining the
variance in the fitted lifetime for different $t'$ bin size, by reducing the $t'$ range
for the fit from 3 to 2 ps and by excluding the lowest $t'$ bin from the fit. 
\begin{table}
\begin{center}
\caption{Contributions to the systematic uncertainty.}
\begin{tabular}{cc}
 \hline
  Contribution &  Uncertainty (fs) \\
 \hline
 \hline
   \cc Momentum            & $ \pm 2$ \\
   two solution bias       & $ \pm 1$ \\
   Split sample            &    0     \\
   $t'$ Resolution         & $ \pm 4$ \\
   Background              & $ \pm 2$ \\
   Fit variant             & $ \pm 8$ \\
 \hline
    Total                  & $ \pm 9$ \\
 \hline
\end{tabular}
\label{table:sys}
\end{center}
\end{table}
Uncertainties in the measurement of particle momenta can lead to a systematic shift in
the reduced proper time. Our studies show this shift to be $4 \pm 2$ fs. 
The final quoted value is adjusted by this amount and a systematic uncertainty of 2 fs
is included in the final systematic uncertainty.
The treatment of the two solution ambiguity in the $\Sigma^+ K^- \pi^+$ mode creates a
small bias in the measured lifetime due to an overestimation of the background in the
signal region. This effect is less than 1 fs on the total lifetime.
Systematic effects due to the background were investigated by varying the width of the
sideband regions, and by altering the background level by imposing a minimum 
separation between the primary and secondary vertices of 1.5 mm. The variance from
these tests is added to the systematic uncertainty.
We tested different fit conditions; excluding MV type decays from the
lifetime fit, taking the weighted average of the split samples, and using a combined
fit to the mass shape and reduced proper time of \skpi as an alternative method of
treating the two solution ambiguity.
The systematic uncertainty due to the variation in fit conditions is taken to be the
sample variance since we consider all of the measurements to be equally valid. The 
systematic contribution is found to be 8 fs. 

\begin{figure*}[t]
  \begin{center}
     \includegraphics[width=2.5in]{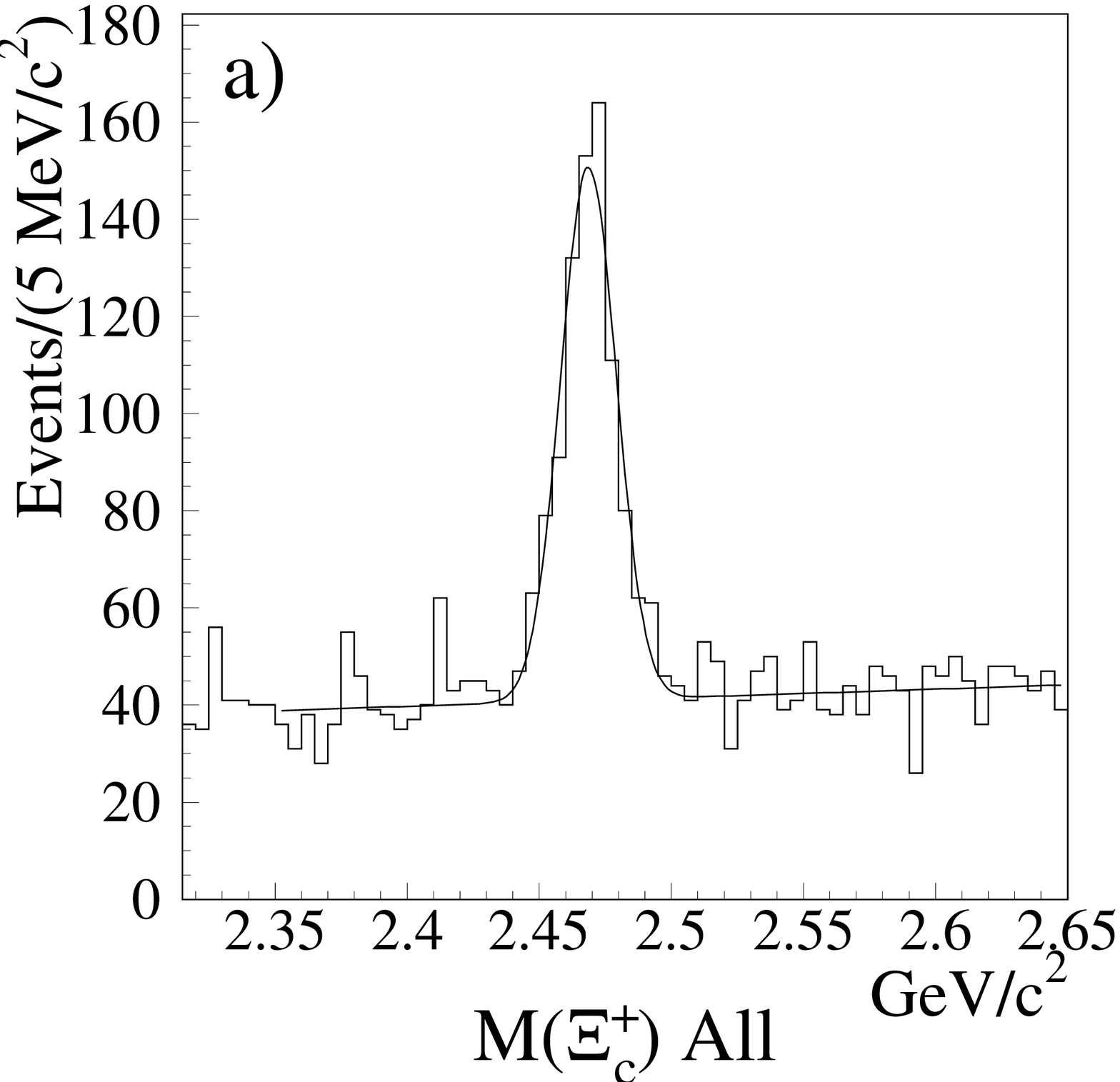}
     \includegraphics[width=2.5in]{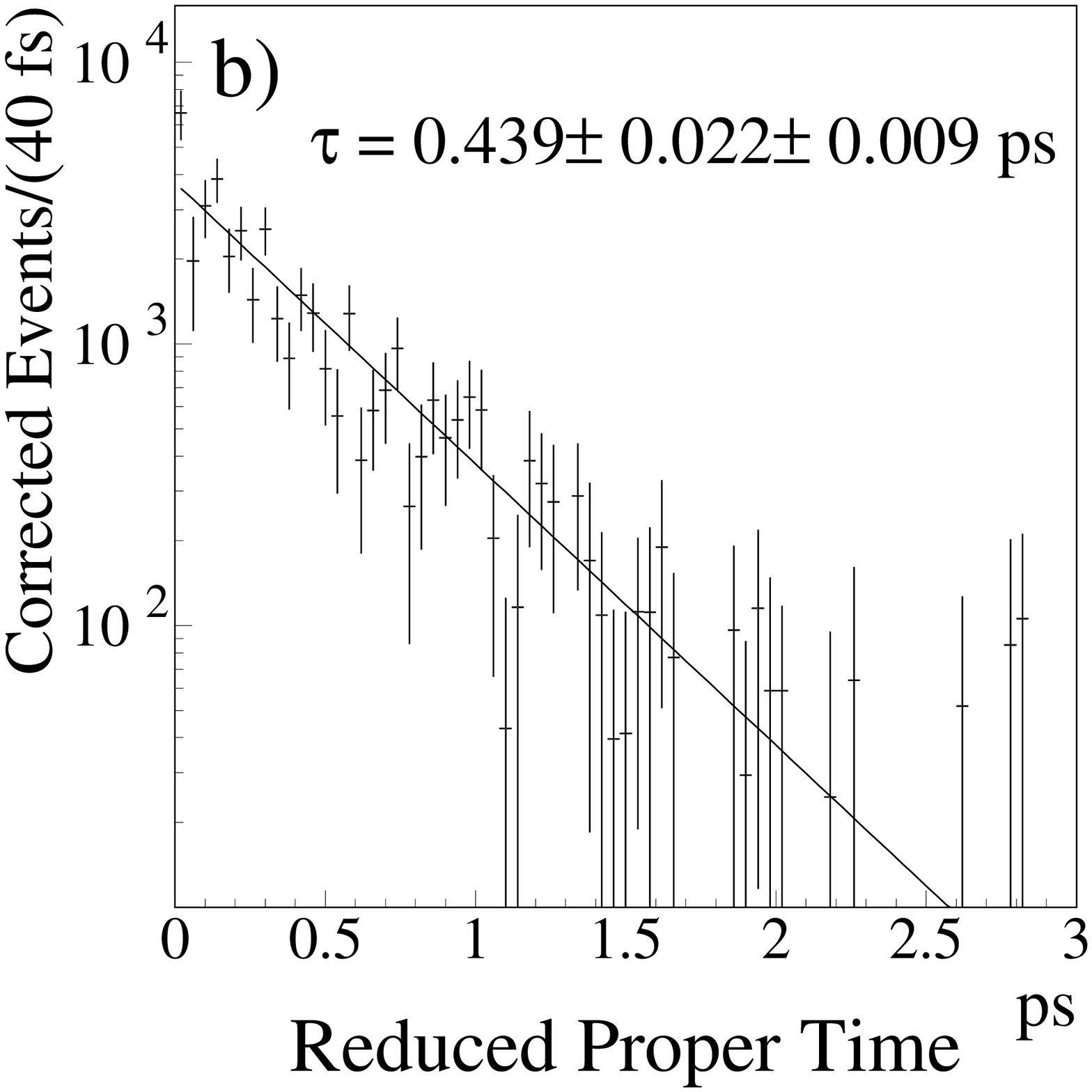}
  \caption{a) The invariant mass for the combined sample. 
b) The combined lifetime fit of the \cc modes with a background
subtracted, Monte Carlo corrected, reduced proper time distribution. 
}
  \label{fig:ltime}
 \end{center}
\end{figure*}

The components of the systematic error are presented in Table \ref{table:sys}.
Adding these contribution in quadrature gives a total systematic uncertainty of 9 fs.
Figure \ref{fig:ltime}(b) shows the background subtracted, Monte Carlo corrected,
reduced proper time distribution for the $\Xi_c^+$ signal. 
\section{Summary}
We have measured the \cc using four decay modes which occur in eight different
topologies with a combined sample of 532.4 $\pm$ 30.4 events. We find the lifetime to
be $0.439 \pm 0.022 \pm 0.009$ ps where the first error is statistical and the second
is systematic.
Our result is compared with other experimental values in Table
\ref{table:ltlist}.
\begin{table}
\begin{center}
\caption{$\Xi_c^+$ lifetime measurements}
\begin{tabular}{cccc}
 \hline
  Experiment &  Lifetime (ps)  &  Events & Year  \\
 \hline
 \hline
   WA62 \cite{Biagi:1985hz}    & $0.48^+_-{}^{0.21+ 0.20}_{0.15 -0.15}$ & 53  & 1985\\
   E400       \cite{Coteus:1987ar}  & $0.40^+_-{}^{0.18}_{0.12}\pm0.10$ & 102 & 1987\\
   ACCMOR (NA32)\cite{Barlag:1989sw}& $0.20^+_-{}^{0.11}_{0.06}$        & 6   & 1989\\
   E687 \cite{Frabetti:1993yy}      & $0.41^+_-{}^{0.11}_{0.08}\pm0.02$ & 30  & 1993\\
   E687 \cite{Frabetti:1998kr}      & $0.34^+_-{}^{0.07}_{0.05}\pm0.02$ & 56  & 1998\\
   PDG (average) \cite{Groom:2000in}& $0.33^+_-{}^{0.06}_{0.04}$        & --  & 2000\\
  This Measurement                  & $0.439\pm 0.022\pm0.009$          & 532 & 2001\\
 \hline
\end{tabular}
\label{table:ltlist}
\end{center}
\end{table}
Our measurement uncertainty is a factor of two better than those of the current
world average \cite{Groom:2000in}.
As discussed in the introduction, theoretical models predict the lifetime hierarchy of the
charmed baryons. A number of authors
\cite{Guberina:1997yx,Bigi:1996sg,Blok:1991st,Cheng:1997xb}
predict that $\tau(\Xi_c^+)>\tau(\Lambda_c^+)$ where the inequality
represents a factor of about 1.3. 
Using the $\Lambda_c^+$ lifetime average of PDG, CLEO, and SELEX
\cite{Groom:2000in,Mahmood:2000tw,Kushnirenko:2000ed} 
($0.1916 \pm 0.0054$ ps)
and the $\Xi_c^+$ lifetime reported in this
paper, a ratio $\tau(\Xi_c^+)/\tau(\Lambda_c^+)= 2.29  \pm 0.14$, is obtained.
Our well measured ratio is significantly different from predictions of order 1.3.

\section{Acknowledgments}

We wish to acknowledge the assistance of the staffs of Fermi National
Accelerator Laboratory, the INFN of Italy, and the physics departments of the
collaborating institutions. This research was supported in part by the U.~S.
National Science Foundation, the U.~S. Department of Energy, the Italian
Istituto Nazionale di Fisica Nucleare and Ministero dell'Universit\`a e della
Ricerca Scientifica e Tecnologica, the Brazilian Conselho Nacional de
Desenvolvimento Cient\'{\i}fico e Tecnol\'ogico, CONACyT-M\'exico, the Korean
Ministry of Education, and the Korean Science and Engineering Foundation.

\bibliographystyle{myapsrev}
\bibliography{xc_life}

\end{document}